# Searching for Extraterrestrial Intelligence by Locating Potential ET Communication Networks in Space

**Ross Davis, PhD**

**Indiana University**

**July 9th, 2019**

For inquiries, please contact:

Ross Davis, PhD

rossdavi@iuk.edu

**Abstract**

There have been periodic efforts in recent decades to search for extraterrestrial intelligence (SETI), especially by trying to find an extraterrestrial (ET) radio signal or other technosignature in space. Yet, no such technosignatures have been found. Considering the vastness of space, finding such technosignatures has been described as trying to find a needle in a cosmic haystack. To help resolve this, two hypotheses are proposed to aid SETI researchers in narrowing the search for ET technosignatures, based on a network analysis approach to locate where in space potential ET communication networks would most likely be. A potential ET communication network can use exoplanets as communication access points (e.g., placing a communication satellite into a planet's orbit, and/or an antenna on a planet's surface). The approach uses a topology where exoplanets are represented as nodes, and the lines of average distance (generalized communication paths) between adjacent exoplanets are represented as edges; the nodes and edges form local and wide planetary networks. Using the approach and data visualization on exoplanet databases can highlight locations of potential ET communication networks in space. The first hypothesis posits that an ET technosignature would more likely appear in a potentially habitable solar system containing a high concentration of planets, wherein the planets function as communication access points to facilitate a potential ET communication network. The second hypothesis posits that an ET technosignature would more likely appear in a highly concentrated cluster of potentially habitable solar systems. Contributions to the SETI field can be increased accuracy in finding ET technosignatures, increased accuracy in reaching a Schelling point (a mutual realization of how we and an ET intelligence can find each other), and promoting interdisciplinary SETI research.

1. **Introduction**

Since around 1959 there have been periodic efforts on Earth to search for extraterrestrial intelligence (SETI) by way of detecting of extraterrestrial (ET) radio signals from outer space—a type of SETI known as communication SETI (Wright, 2018). So far though, no such signals have been reliably found. Considering the vastness of space and the countless stars in it, trying to find an ET radio signal has been described as being similar to trying to find a needle in a cosmic haystack. Still, a number of researchers engaged in communication SETI remain undaunted as they pursue innovative ways in which to detect ET transmission signals coming from space, such as going beyond the radio spectrum to explore optical (think lasers) and infrared. More broadly conceived, ET radio, optical, and infrared space signals, as well as other technological, non-biological signs of ET intelligence are collectively known as technosignatures (Wright et al., 2018).

To aid SETI researchers in narrowing the search for ET technosignatures, this paper proposes two hypotheses based on a network analysis approach to locate where in space potential ET communication networks would most likely be. These hypothetical networks would be capable of exhibiting ET technosignatures that include transmitted radio, optical, or other types of electromagnetic signals. In such an approach, a potential ET communication network has a physical topology wherein the planets that a communicative ET intelligence could use as communication access points (in its own solar system and beyond) represent nodes, and the lines of average distance (generalized communication paths) between such planets represent edges. A planet can serve as communication access point in various ways, such as by placing a communication satellite into orbit around the planet, or by stationing an antenna on the planet's surface.

The following scenario, or algorithm, illustrates how the hypotheses can be tested by applying data visualization in conjunction with network analysis of exoplanet databases (such as those developed from NASA's Kepler and TESS space missions); the hypotheses posited to be situated where ET communication networks would most likely be.

- A. An astronomical chart, preferably 3-dimensional (3D), would be generated showing local planetary networks (LPNs) and wide planetary networks (WPNs). Each LPN consists of exoplanets within a potentially habitable solar system, where each exoplanet represents a node, and the line of the average distance between each exoplanet represents an edge. (An would be a generalized communication path if there was any potential ET communication network.) Encompassing the LPNs would be the broader WPNs. Each WPN consists of clusters of potentially habitable solar systems, where each of these solar systems represents a node, and the line of the average distance between each solar system represents an edge.

- B. The graphical depiction of the nodes (exoplanets) in the chart, such as their size and color, and the graphical depiction of the edges, such as their thickness and color, would be determined by planetary-scale measures associated with the extent to which conditions in space facilitate a potential ET communication network. A planetary-scale measure as such that can determine the graphical depiction of nodes can include, though not necessarily be limited to, a planet's diameter, mass, or even a composite/index measure. And a planetary-scale measure that can determine the graphical depiction of edges can include the degree of space dust between each neighboring exoplanet, the orbital eccentricities of each planet around its host star(s), or a composite/index measure.

C. After applying the aforementioned data visualization and network analysis, the LPNs and/or WPNs where potential ET communications networks would most likely be (based on the planetary-scale measures for nodes and edges previously described) would graphically stand out in the 3D astronomical chart via highlighted nodes and edges.

A simplified example of this in terms of network analysis would be to consider three neighboring exoplanets called Alpha, Beta, and Gamma in a potentially habitable solar system. Consider that exoplanet Alpha would be a gas giant like Jupiter, while exoplanets Beta and Gamma would be solid Earth-like planets. If we graphically depicted solid Earth-like planets as nodes represented by solid black circles, and gas giants as nodes represented by hollow black circles, then Alpha, Beta, and Gamma would be depicted accordingly. Now consider that there would be an observable dust cloud or similar barrier between Alpha and Beta (but not between Beta and Gamma), which could impair the ability of a potential ET communication network to span between Alpha and Beta, particularly where such as network entailed radio transmissions. If we graphically depicted edges in bold between neighboring nodes for when there would be no dust cloud or similar communication barrier in between the nodes, and unbold edges for non-barrier conditions, then the edges between the nodes (representing Alpha, Beta, and Gamma) would be depicted accordingly. In sum, in this simplified example, planetary networks depicted mainly by solid black circles connected with bold edges can highlight where potential ET communication networks would likely be.

Considering the aforementioned discussion, the two hypothesis are as follows:

    H1:    An ET technosignature would more likely appear in a potentially habitable solar system containing a high concentration of exoplanets, where the exoplanets function as communication access points to facilitate a potential ET communication network.

    H2:    An ET technosignature would more likely appear in a highly concentrated cluster of potentially habitable solar systems, wherein the potentially habitable solar systems (by way of their respective exoplanets) function as communication access points to facilitate a potential ET communication network.

The hypotheses have been inferred from an interdisciplinary approach incorporating concepts from information science, the social sciences, and the physical sciences. These concepts are elaborated on in the following section.

## 2. Concepts

The hypotheses incorporate the concepts of communicative ET intelligence, potentially habitable solar systems, and space-based ET communications networks. Each of these concepts is described as follows, to provide context relative to the hypotheses.

*2.1. Communicative ET Intelligence*

A communicative ET intelligence is a species "not originating recently on Earth" that transmits technosignatures through space consisting of information on carriers such as photons (Wright, 2018; Wright et al., 2018). A communicative ET intelligence would be represented by a factor in the renowned Drake Equation, which is (Drake, 1965):

$$N = R_* \cdot f_p \cdot n_e \cdot f_l \cdot f_i \cdot f_c \cdot L$$

In the equation, $N$ represents the number of communicative intelligent civilizations in the Milky Way galaxy that are detectable. $R_*$ represents the average rate of star formation in the Milky Way. $f_p$ is the fraction of stars in the Milky Way that have planets. $n_e$ is the average number of planets in the Milky Way that support life for each star that has planets. $f_l$ represents the fraction of planets in the Milky Way that can support life and where life would actually develop. $f_i$ represents the fraction of planets in the Milky Way that have life that develops into intelligent life. $f_c$ is the fraction such intelligent life that transmits detectable signals into space. And $L$ is the length of time for which such intelligent life transmits detectable signals into space.

Considering the virtually ubiquitous evolutionary processes on Earth, such as the biological (Darwin, 1876), sociological/anthropological (Nolan and Lenski, 2015), and informational (Vigo, 2013), it is conceivable that similar evolutionary processes can occur on a potentially habitable exoplanet such that a communicative ET intelligence can emerge (Grinspoon, 2018), notwithstanding that it would not necessarily be in lockstep with our evolution. It can be as though evolution in general is a cosmic universal, analogous to a cultural universal such as music found throughout several regional human societies here on Earth (Brown, 1991).

*2.2. Potentially Habitable Solar Systems*

A potentially habitable solar system, beyond Earth's own solar system, is a solar system that contains at least one exoplanet potentially located in the solar system's habitable zone. The habitable zone is a circumstellar region within a solar system wherein a terrestrial mass planet, having between 0.1 and 10 Earth masses and favorable atmospheric conditions, can sustain liquid water on its surface (Huang 1959; Hart 1978; Kasting et al. 1993; Underwood et al. 2003; Selsis et al. 2007; Kaltenegger & Sasselov 2011; Kopparapu et al. 2013). Such a planet can support life ranging from microorganisms to plants and animals (one reference for determining potentially habitable solar systems involving exoplanets is the Habitable Exoplanets Catalog maintained by the Planetary Habitability Laboratory at the University of Puerto Rico at Arecibo). A potentially habitable solar system does not mean that life necessarily exists there, rather it has the potential to exist (Planetary Habitability Laboratory, 2018). There can be an incredible opportunity to discover potentially habitable solar systems beyond Earth, considering that there can be as many as 11 billion Earth-size exoplanets orbiting the habitable zone of Sun-like stars in the Milky Way galaxy (Patigura et al., 2013).

*2.3. Hypothetical ET Communication Networks in Space*

The network analysis approach in conjunction with the hypotheses are based on the following idea: The limits of intelligent communication across space can be captured by supposing the existence of communication networks. While we are only familiar with our own modern space-based communication network extending from Earth, we can draw upon its characteristics as a basic analogy to aid us in locating where in space a potential ET communication would most likely be. The physical topology of our space-based communication network effectively uses the planets in our own solar system as nodes, and the lines of average distance (mean communication paths) between adjacent planets as edges. For instance, consider the planet Jupiter having served as node to help facilitate NASA's space-based communication network. This significantly took shape in 1973 when NASA's Pioneer 10 space probe reached Jupiter. Upon reaching Jupiter, Pioneer 10 carried out a multitude of experiments ranging from studying interplanetary and planetary magnetic fields, solar wind parameters, cosmic rays, to the atmosphere of Jupiter and some of its satellites. Pioneer transmitted data from these experiments back to Earth, in the megabytes range. This contributed to our ability to use this information, and Jupiter's gravitational pull, to network further on to Saturn, Uranus, and Neptune with the subsequent Voyager 1 and 2 missions several years later (NASA, 2018). Moreover, by having a higher concentration of planets in our solar system (8 planets) than we otherwise could have had (for example, if we only had 3 planets), conditions were more favorable for us to facilitate a communication network in space.

    The second hypothesis pertaining to the likelihood of finding ET technosignatures in a highly concentrated cluster of potentially habitable solar systems considers the following: If a communicative ET intelligence has the capability to expand its space-based communications network beyond its own solar system to another one, it can do so more conveniently if the other solar system was close by instead of far away, especially if there would be considerations such as potential attenuation of high frequency components of an ET transmission technosignature from interstellar scattering between solar systems. It "will require special transmission schemes if broadcasts are made over long distances in the galactic plane" (Shostak, 1995). This is with respect to Shannon's Law---$C = W \cdot \log_2(1 + P/N)$, which determines "the amount of

information C (bits/second) that can be conveyed with a channel of bandwidth W" (Shannon, 1948).

Each hypothesis by itself is deemed as an essential, yet not sufficient condition in the likelihood of detecting a technosignature of a communicative ET intelligence.

**Conclusions**

Contributions to the SETI field can be 1) increased accuracy in finding ET technosignatures through an enhanced search algorithm of exoplanet databases using a network analysis approach, 2) increased accuracy in reaching a Schelling Point, and 3) promoting interdisciplinary SETI research.

Regarding increased accuracy in finding ET technosignatures through an enhanced search algorithm of exoplanet databases using a network analysis approach, this would be especially advantageous in fine-tuning targeted sky searches (besides all-sky searches) in terms of narrowing the search area. This can save costs in terms of time and the amount of data to analyze. And a network analysis approach can offer enhanced data visualization of exoplanet data, especially on the level of big data (for examples of network analysis of other kinds of data, refer to Nykamp, 2019). An example could be a 3D astronomical chart showing several highly concentrated clusters of potentially habitable solar systems. The solar systems in each cluster would be depicted as nodes, and the distance lines (a communication path) between each solar system depicted as edges. The nodes and edges in each cluster would represent a potential ET communication networks. The chart can also show potentially habitable solar systems not considered part of a cluster, yet where these solar systems would still have a high concentration of planets.

 As far as increased accuracy in reaching a Schelling Point, this is the point at which we and a communicative ET intelligence both happen to come to the same realization of how we both should try to discover each other, prior to first contact.  An example is a communicative ET intelligence setting up a beacon to transmit a radio signal on a "magic frequency"—the frequency that the ET intelligence hopes we will dial into; the signal may say something as simple as "We are here!" (Wright, 2018).  Besides searching for the magic frequency, SETI researchers can draw upon the hypotheses to also consider where such a frequency may likely originate from in the cosmos.  This can help narrow the search for such a frequency, thus saving time, money, and other resources in the process.

 Since the hypotheses draw upon concepts from an interdisciplinary perspective (such as network analysis used in information science, the social sciences, and the physical sciences), it can promote interdisciplinary research.

 A limitation of the hypotheses is that their rationale would mainly apply to communicative ET intelligences whose evolution would be basically comparable with ours here on Earth.  Thus, this may be interpreted as being anthropocentric to an extent.  Yet the approach does not presume that the evolution of a communicative ET intelligence would necessarily parallel ours in every aspect.  Nevertheless, it is conceivable that among the estimated 11 billion Earth-like exoplanets that can be in our galaxy, there can be a chance for a communicative ET intelligence whose evolution is comparable to ours.  In other words, there can be a range of diversity among communicative ET intelligences,

**Acknowledgements**

Thanks to the Lunar and Planetary Institute, Universities Space Research Association, and NASA for organizing the 2018 NASA Technosignatures Workshop that helped to inspire this paper. Thanks to Seth Shostak of the SETI Institute, David Grinspoon of the Planetary Science Institute, and Kathryn Denning of York University for suggesting sources; and Ronaldo Vigo of Ohio University for providing editorial assistance. And thanks to other SETI colleagues who responded to related scientific questions via the International Academy of Astronautics (IAA) SETI Committee's Google group forum.

**Author Disclosure Statement**

No competing financial interests exist.

**References**

Brown, D.E. (1991) *Human Universals*. Temple University Press, Philadelphia.

Darwin, C.R. (1876) *The Origin of Species by Means of Natural Selection, or the Preservation of Favoured Races in the Struggle for Life*, 6$^{th}$ edition. John Murray, London.

Drake, F. (1965) The radio search for intelligent extraterrestrial life. In *Current Aspects of Exobiology*, edited by G. Mamikunian and M.H. Briggs, Oxford University Press, pp. 323-45.

Grinspoon, D. (2018) Cognitive planetary transitions: An astrobiological perspective on the "Sapiezoic Eon". Presented at the 2018 NASA Technosignatures Workshop, Lunar and Planetary Institute, Houston, TX. Available online at

SETI by Locating Potential ET Communication Networks
https://www.hou.usra.edu/meetings/technosignatures2018/presentation/?video=grinspoon.mp4

Hart, M.H. (1978) The evolution of the atmosphere of the earth. In *Icarus*, 33: 23-39.

Huang, S.S. (1959) Occurrence of life outside the solar system. In *American Scientist*, 47, 397.

Kaltenegger, L. and Sasselov, D. (2011) Exploring the habitable zone for Kepler planetary candidates. In *The Astrophysical Journal Letters*, 736, 2, L25. Available online at http://iopscience.iop.org/article/10.1088/2041-8205/736/2/L25/pdf

Kasting, J.F., Whitmire, D.P., and Reynolds, R.T. (1993) Habitable zones around main sequence stars. In *Icarus*, 101, 1: 108-228.

Kopparapu, R.K., Ramirez, R., Kasting, J.F. Eymet, V., Robinson, T.D., Mahadevan, S., Terrien, R.C., Domagal-Goldman, S., Meadows, V., and Deshpande, R. (2013) Habitable zones around main-sequence stars: New estimates. In *The Astrophyical Journal*, 765, 131. Available online at http://iopscience.iop.org/article/10.1088/0004-637X/765/2/131/pdf

NASA. (2018) Pioneer 10. NASA Space Science Data Coordinated Archive (NSSDCA). Available online at https://nssdc.gsfc.nasa.gov/nmc/spacecraft/display.action?id=1972-012A

Nolan, P. and Lenski, G. (2015) *Human Societies: An Introduction to Macrosociology*. Oxford University Press, New York, NY.

Nykamp D.Q. (2019) "An introduction to networks." Math Insight. Available online at http://mathinsight.org/network_introduction

Patigura, E.A., Howard, A.W., and Marcy, G.W. (2013) Prevalence of Earth-size planets orbiting Sun-like stars. Proceedings of the National Academy of Sciences, November 2013. Available online at https://www.pnas.org/content/pnas/110/48/19273.full.pdf

Planetary Habitability Laboratory, University of Puerto Rico at Arecibo. (2018) HEC: About the Habitable Exoplanets Catalog. Available online at http://phl.upr.edu/projects/habitable-exoplanets-catalog/about

Selsis, F., Kasting, J.F., Levrard, B., Paillet, J., Ribas, I., and Delfosse, X. (2007) Habitable planets around the star Gliese 581? In *Astronomy and Astrophysics*, 476, 3: 1373-1387. Available online at https://www.aanda.org/articles/aa/pdf/2007/48/aa8091-07.pdf

Shannon, C.E. (1948) A mathematical theory of communication. In *Bell System Technical Journal*, 27: 379-423, 623-656.

Shostak, G.S. (1995) SETI at wider bandwidths? In *Progress in the Search of Extraterrestrial Life*, ASP Conference Series, 74, ed. G.S. Shostak: pp. 447-454. Astronomical Society of the Pacific, San Francisco.

Underwood, D.R., Jones, B.W., and Sleep, P.N. (2003) The evolution of habitable zones during stellar lifetimes and its implications on the search for extraterrestrial life. In *International Journal of Astrobiology*, 2, 4: 289-299.

Vigo, R. (2013) Complexity over uncertainty in generalized representational information theory (GRIT): A structure-sensitive general theory of information. In *Information*, 4: 1-30. Available online at https://www.mdpi.com/2078-2489/4/1/1/pdf

Wright, J.T. (2018) Taxonomy and jargon in SETI as an interdisciplinary field of study. From the 2018 Decoding Alien Intelligence Workshop, SETI Institute, Mountain View, CA. Available online at https://daiworkshop.seti.org/sites/default/files/workshop-2018/Wright%20-%20Taxonomy%20and%20Jargon%20in%20SETI%20as%20an%20Interdisciplinary%20Field%20of%20Study.pdf

Wright, J.T., Sheikh, S., Almár, I., Denning, K., Dick, S., and Tarter, J. (2018) Recommendations from the ad hoc committee on SETI nomenclature. Available online at https://arxiv.org/ftp/arxiv/papers/1809/1809.06857.pdf